\documentclass{emulateapj}

\newcommand{\noprint}[1]{}

\newcommand{\myemail}{buenzli@mpia.de}

\slugcomment{Draft version \today}

\shorttitle{Cloud structure of the nearest brown dwarfs}
\shortauthors{Buenzli et al.}

\begin{document}

\title{Cloud structure of the nearest brown dwarfs: Spectroscopic variability \\ of Luhman\,16AB from the \it{Hubble Space Telescope}}

\author{Esther Buenzli$^{1}$, Didier Saumon$^{2}$, Mark S. Marley$^{3}$, D\'aniel Apai$^{4,5}$, Jacqueline Radigan$^6$, \\ Luigi R. Bedin$^7$, I. Neill Reid$^6$,  and Caroline V. Morley$^8$} 

\affil{$^1$Max Planck Institute for Astronomy, K\"onigstuhl 17, D-69117 Heidelberg, Germany, \myemail}
\affil{$^2$Los Alamos National Laboratory, Mail Stop F663, Los Alamos, NM 87545, USA}
\affil{$^3$NASA Ames Research Center, MS-245-3, Moffett Field, CA 94035, USA}
\affil{$^4$Department of Astronomy, University of Arizona, 933 N. Cherry Avenue, Tucson, AZ 85721, USA}
\affil{$^5$Department of Planetary Sciences, University of Arizona, 1629 E. University Blvd, Tucson AZ 85721, USA}
\affil{$^6$Space Telescope Science Institute, 3700 San Martin Drive, Baltimore, MD 21218, USA}
\affil{$^7$INAF - Osservatorio Astronomico di Padova, Vicolo dell'Osservatorio 5, I-35122 Padova, Italy}
\affil{$^8$Department of Astronomy and Astrophysics, University of California, Santa Cruz, CA 95064, USA}

\begin{abstract}
The binary brown dwarf WISE\,J104915.57$-$531906.1 (also Luhman\,16AB), composed of a late L and early T dwarf, is a prototypical L/T transition flux reversal binary located at only 2\,pc distance. Luhman\,16B is a known variable whose light curves evolve rapidly. We present spatially resolved spectroscopic time-series of Luhman\,16A and B covering 6.5\,h using HST/WFC3 at 1.1 to 1.66\,$\mu$m. The small, count-dependent variability of Luhman\,16A at the beginning of the observations likely stems from instrumental systematics; Luhman\,16A appears non-variable above $\approx$0.4\%. Its spectrum is well fit by a single cloud layer with intermediate cloud thickness ($f_{\mathrm{sed}}=2$, T$_{\mathrm{eff}}=1200$\,K). Luhman\,16B varies at all wavelengths with peak-to-valley amplitudes of 7-11\%. The amplitude and light curve shape changes over only one rotation period. The lowest relative amplitude is found in the deep water absorption band at 1.4 $\mu$m, otherwise it mostly decreases gradually from the blue to the red edge of the spectrum. This is very similar to the other two known highly variable early T dwarfs.  A two-component cloud model accounts for most of the variability, although small deviations are seen in the water absorption band. We fit the mean spectrum and relative amplitudes with a linear combination of two models of a warm, thinner cloud (T$_{\mathrm{eff}}=1300$\,K, $f_{\mathrm{sed}}=3$) and a cooler, thicker cloud (T$_{\mathrm{eff}}=1000-1100$\,K, $f_{\mathrm{sed}}=1$), assuming out-of-equilibrium atmospheric chemistry. A cloud as for Luhman\,16A but with holes cannot reproduce the variability of Luhman\,16B, indicating more complex cloud evolution through the L/T transition. The projected separation of the binary has decreased by $\approx0\farcs3$ in 8 months. 
\end{abstract}


\section{Introduction}

\begin{figure*}
\epsscale{0.8}
\centering
\includegraphics*{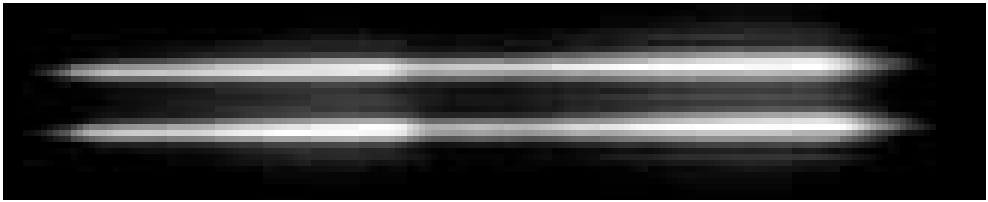}
\caption{1st order spectra of Luhman\,16A (upper) and B (lower) recorded on the detector, spanning from about 1.05 to 1.75 $\mu$m. The dispersion direction is approximately toward North-East. The gray scale is logarithmic.
\label{fig:rawspec}}
\end{figure*}

The recently discovered brown dwarf binary WISE\,J104915.57$-$531906.1 \citep{luhman13}, hereafter Luhman\,16AB, is one of only a few known spatially resolved binaries where both components are located at the transition of L to T spectral type \citep[L7.5 and T0.5,][]{kniazev13, burgasser13}. Its very large parallax of $495\pm4.6$ mas \citep{boffin14} puts it at a distance of only $2.020\pm0.019$ pc, making it the third closest known system from the Sun after $\alpha$ Cen and Barnard's star. Because of that proximity, they are by far the brightest brown dwarfs of their spectral type, allowing  studies not previously possible for fainter objects. For example, \citet{crossfield14} were able to use time-resolved very high resolution spectroscopy to create a surface map by applying Doppler imaging for the first time to a brown dwarf. The map showed a heterogeneous surface structure for Luhman\,16B that may be linked to patchy cloud cover. 

Heterogeneous cloud cover has already been inferred for other early T dwarfs \citep[e.g.][]{artigau09, radigan12, apai13, radigan14} as well as for Luhman\,16B \citep{gillon13,biller13,burgasser14} from measurements of photometric and/or spectroscopic variability. The removal of cloud opacity is the dominant driver of the spectral changes happening between late L to mid T type dwarfs \citep[e.g.][]{kirkpatrick05, cushing08, stephens09}. The change in near-infrared colors from red to blue and the brightening of the J band suggest that the clouds, likely composed of silicates and iron, that form opaque layers in late L dwarfs are fully removed from the visible photosphere by spectral type of $\approx$T5 \citep{dupuy12}. These changes happen at nearly constant effective temperatures of T$_{\mathrm{eff}}\approx1,300$\,K. 

One mechanism proposed to explain the color evolution, the re-emergence of the FeH feature from early to mid T dwarfs \citep{burgasser02,burgasser03}, and the photometric variability is the appearance and growth of holes in the clouds \citep{ackerman01, burgasser02, marley10}, where flux can then emerge from deeper, hotter regions. Alternatively, cloud thinning through growth of particle size, increased sedimentation efficiency and rapid rain out may also remove the clouds \citep{tsuji04, knapp04, burrows06, saumon08}. The multi-wavelength variability of the two early T dwarfs 2MASS\,J21392676+0220226 (hereafter 2M2139) and SIMP\,J013656.57+093347.3 (hereafter SIMP0136) are not compatible with fully cleared holes in clouds. Instead, models that are a combination of varying covering fraction of thin and thick clouds over one rotation period can reproduce many of the characteristics of the variability  \citep{radigan12,apai13,buenzli14b}. 

With Luhman\,16B, a third early T dwarf with multiple percent variability is now available to test whether the two-component thin/thick cloud model can represent the cloud structure at the beginning of the L/T transition. The Luhman\,16 system offers many additional benefits to significantly increase our understanding of cloud structure at the L/T transition. Its brightness offers the possibility for studies over a very broad wavelength range, even in the optical \citep{biller13}, and a very good signal to noise ratio at intermediate \citep{faherty14} or very high resolution \citep{crossfield14}. The distance is already known very precisely and the orbital motion will eventually lead to a measurement of the dynamical mass; the orbital time scale is $\approx25$~years at a separation of $\approx$3~AU. Finally, the A component is most likely co-eval with the B component, implying that their age and metallicity are equal. The A component is brighter overall, but fainter in the Y and J bands \citep{burgasser13}. This flux reversal suggests that cloud evolution has progressed less far than for the B component and it therefore provides a comparison point before the onset of the L/T transition. 

Many characteristics of the Luhman\,16AB binary have already been constrained. The discovery of lithium in both objects points to an age between 0.1 and 3~Gyr \citep{faherty14}. Their masses are therefore in the range of 20-65~M$_{J}$. From their bolometric luminosities, their effective temperatures are very similar with $T_{\mathrm{eff}} = 1310\pm30$\,K for A and $1280\pm75$\,K for B, while the brightness temperature is about 50 K higher for B in regions that are dominated by condensate grain scattering \citep{faherty14}. The $v \sin{i}$ measurements indicate that the inclination of Luhman\,16B is $<30^\circ$ from equator-on \citep{crossfield14}. The rotation period has been determined as close to 5 hours, with formal measurements $4.87\pm0.01$ \citep{gillon13} or $5.05\pm0.1$~h \citep{burgasser14}. Precise determination of the rotation period is difficult because the light curve shape evolves on time scales of only one rotation period. 

In this paper, we present spatially resolved spectroscopic time series obtained with the Hubble Space Telescope (HST) for both Luhman\,16 A and B. This allows a study of their variability amplitudes independently and with high precision, and gives access to features such as the 1.4~$\mu$m water band that are not obtainable from the ground but are an important tracer of the vertical extension of the cloud. In Section 2 we describe the observations and data reduction. In Section 3 we present and analyze the measured spectroscopic variability and its evolution over one rotation period. We also provide an update on the binary separation. In Section 4 we model the observations with patchy cloud models and in Section 5 we discuss and compare to previous results. Our conclusions are presented in Section 6. 

\section{Observations and data reduction}

\subsection{Observations}

\begin{figure*}
\epsscale{1.15}
\plotone{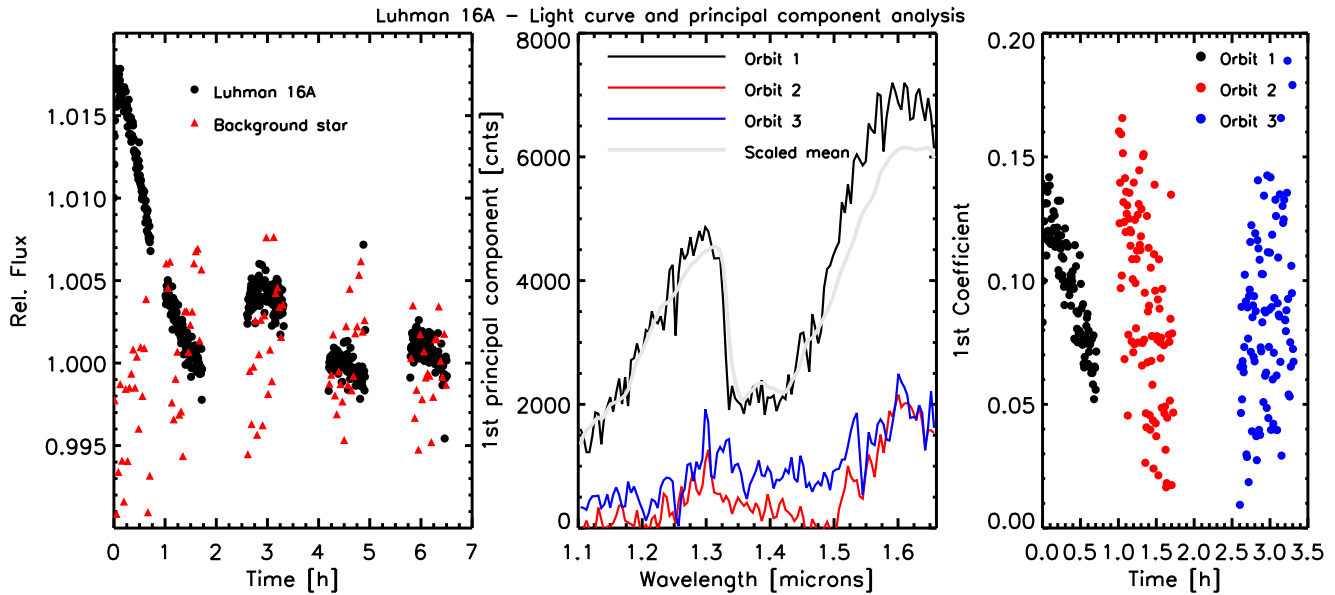}
\caption{Observations and principal component analysis for Luhman\,16A. Left: Integrated raw light curve for Luhman\,16 A (black dots) and a fainter background star (red triangles, binned by factor of 5). Middle: First principal component (see text) derived for orbits 1 (black), 2 (red) and 3 (blue). The average recorded flux in counts (scaled) is shown in gray. The first component for the first orbit appears to be largely proportional to the observed count rate. Right: Coefficients corresponding to the first principal component. 
\label{fig:luh16a}}
\end{figure*}

We observed Luhman\,16AB on 08 November 2013 between 06:34 and 13:06 (UT)  with HST/WFC3 in the infrared channel (Program \#13280). We observed for 5 consecutive orbits with gaps in between
because the target is hidden behind the Earth for parts of the orbit. The effective observation length per orbit was shortened from a maximum visibility of 57~min to 44~min because of acquisition and because the maximum number of files that can be stored in the WFC3 buffer is only 304 (counting each non-destructive read). This limit was reached before the end of the target visibility, and the necessary buffer dump took up the rest of the available time per orbit, in which time no further observations can be taken. The gap between the first and second orbit was shorter than between the other orbits because the target is located relatively close to the HST continuous viewing zone. In that case, the observations of the target within the first orbit of a multi-orbit visit can be pushed to as late in the visibility as possible to optimize the schedule.

The detector is a Teledyne HgCdTe with a size of $1024 \times 1024$ pixels, but we used the $256 \times 256$ subarray mode because the full array would allow even fewer observations
to be stored before a buffer dump. The pixel size is $\approx$0\farcs13, resulting in an field of view of $\approx30\arcsec \times 30$\arcsec. 
At the beginning of each orbit we acquired a direct image through the F132N filter used to measure the location of the sources on the detector for precise wavelength calibration. For the remainder of the orbit, we used the G141 grism to take spectral time series of the binary. The first order of the spectra is fully captured on the subarray, while the zeroth and second
orders are not recorded. The spectra have a dispersion of 4.65 nm pixel$^{-1}$ and span $\approx$140 pixels. The orientation of the space craft was set to have the line connecting the binary (nearly) perpendicular to the grism dispersion direction to minimize overlap of the two spectra (Fig.\,\ref{fig:rawspec}). A number of background objects are also visible in the field, none of which overlap with the brown dwarfs. One is the full first order spectrum of a fainter background star, several are very faint and are likely higher order spectra of more distant background stars. A zero-order image is also visible in the field. 

We used the SPARS10 readout mode because the minimum exposure time of the SPARS25 mode, previously used successfully for similar observations of brown dwarfs \citep{buenzli12,apai13,buenzli14} would have resulted in saturation of the target. For each exposure, 3 non-destructive reads of 0 s, 0.278 s and 7.346 s were taken (NSAMP=2) for a total exposure time of 7.624 s. For the direct images, exposure times were only 0.278 s (NSAMP=1). The maximum number of counts recorded in an image was $\approx$25,000. This is well below half-well ($\approx$40,000 counts) where image persistence can become relevant. In each orbit we took 100 exposures with the G141 grism. The cadence (exposure time plus overhead) was 26~s, the length of the spectroscopic time series was 43 min per orbit spanning a total of 6.5 h. 

We observed in staring mode without dithering to avoid errors from pixel-to-pixel sensitivity variations that cannot be corrected to sufficient precision by flatfielding. 

\subsection{Data reduction}

Data reduction was performed using the same method as in \citet{apai13} and \citet{buenzli14}. It is a combination of the standard WFC3 pipeline, custom IDL routines and 
the PyRAF software package aXe\footnote{http://axe-info.stsci.edu} which is used for extracting and calibrating slitless spectroscopic data.  

The WFC3 pipeline \texttt{calfw3} outputs \texttt{flt} files which are the combined images from the non-destructive subreads of an exposure. The pipeline subtracted the zero-read and dark current, flagged bad pixels and corrected for non-linearity and gain. We then identified cosmic rays as $>5\sigma$ outliers for a given pixel compared to the same pixel in the nearest 8 frames in the time series and
replaced them by the median value of the other frames. To correct flagged bad pixels, we interpolated over nearest neighbors in the same row. We only corrected pixels with flag numbers 
4 (dead pixel), 32 (unstable pixel), 256 (saturated) and 512 (bad in flatfield), as the other flagged pixels did not appear to have altered pixel flux above the noise level. 

Before running aXe, we embedded the frames into full-frame images and flagged the extra pixels with a data quality flag to exclude them from processing. We first used the \texttt{axeprep} routine that subtracts the background by scaling a master sky frame. We then ran the \texttt{axecore} routine that flatfields the frames, performs wavelength calibration, extracts the two-dimensional spectra and flux-calibrates with the G141 sensitivity curve. We chose the extraction width to be 7 pixels, which includes 88-92\% of the flux of each binary without notable contamination by the other object (cf. Sect.\,\ref{sec:luh16a}). We apply aperture correction using the values provided by \citet{kuntschner11}, using spline interpolation for intermediate wavelengths. We only consider wavelengths between 1.10 and 1.66 $\mu$m, outside of these the grism sensitivity drops and the larger errors can negatively impact our subsequent principal component analysis. 

\section{Results}

The spectral time series reveal variations for Luhman\,16B at the several percent level at all wavelengths consistent with earlier variability measurements of this object. They are presented and analyzed in detail in Sect.\,\ref{sec:luh16b}. For Luhman\,16A, the strongest variation found is a 1.5\% drop in the first two orbits. Our analysis presented in Sect.\,\ref{sec:luh16a} suggests that this is not intrinsic variability of the brown dwarf but an instrumental systematic. Luhman\,16A is most likely non-variable above our noise level of $\pm0.2-0.4\%$ level at all covered wavelengths. 

\subsection{Luhman\,16A}
\label{sec:luh16a}

The left panel of Fig.\,\ref{fig:luh16a} shows the light curve for Luhman\,16A obtained by integrating over the full spectrum. We exclude the few obvious outliers from further analysis. We find a brief sharp increase over the first 5 exposures and then a strong flux decrease over the first orbit that continues into the second orbit before leveling out. In the third orbit, we find a flux level that is slightly increased by $\approx$0.5\%, and finally a nearly flat curve in the fourth and fifth orbit at the same level as at the end of the second orbit. The shape of the light curve is very different than what would commonly be expected for a rotating brown dwarf with patchy cloud cover and is likely caused by instrumental variations as we demonstrate below. 

WFC3 is known to have strong systematics introduced in particular in the first orbit. In previous brown dwarf variability observations, which were all taken with the SPARS25 mode and longer integration time, a ramp with increasing flux level was found \citep{buenzli12, apai13, buenzli14} that appeared to be largely independent of the count rate. However, observations taken in the SPARS10 mode show a very different behavior that appears to be less stable and shows some indication of a count rate dependence. Transiting planet observations in this mode have revealed a decreasing ramp in the first orbit, though at lower level than what is measured here \citep{mandell13}. In addition, ramps or hooks (continuous or only short flux increase) connected to buffer dumps can be present with strength depending on the array size and number of reads \citep{swain13}. 

To compare systematic effects we retrieve the light curve of the brightest background star, which is still a factor of $\approx$10-50 fainter than the brown dwarfs, even in their water absorption bands. Contrary to Luhman\,16A, we find a slightly increasing ramp in the first orbit (see Fig.\,\ref{fig:luh16a}). This supports the strong count rate dependence of this effect, but implies that the background star cannot be used to correct the light curves of the brown dwarfs. 

We apply a principal component analysis (PCA) in order to determine the spectral characteristics of the changes in the first three orbits with respect to the more constant fourth and fifth orbit. We perform the analysis separately for each orbit in order to avoid the first orbit dominating due to the largest changes. During the PCA, instead of subtracting the mean of the full time series or the individual orbit, we subtract the mean of the counts over the fourth and fifth orbit that we consider to be the baseline. For all three orbits we find that the first component accounts for most of the variations ($>60\%$ for the first orbit and $\approx17\%$ for the second and third, where the variations are not much larger than the noise level). The first principal components and their coefficients are shown in Fig.\,\ref{fig:luh16a}. The results can be interpreted such that $F(\lambda,t)_{\mathrm{orbit1(2,3)}} \approx <F(\lambda,t)_{\mathrm{orbit4,5}}>_t + a_1(t) e_1(\lambda)$, where $a_1(t)$ is the first coefficient and $e_1(\lambda)$ the first principal component (eigenvector), and $F(\lambda,t)$ the flux measured in counts. 

The first principal component of the first orbit reveals that the changes are indeed nearly proportional to the measured count rate, except slightly stronger at wavelengths $>1.5$~$\mu$m. In the second orbit, the count rate dependence is less significant but still present, although most of the variation stems from wavelengths $>1.5$~$\mu$m. In the third orbit, we also find an increase in the flux peaks in addition to an increasing change with wavelength. The origin of these systematic errors remains unclear. For orbits 2 and 3, there may be some influence from drifts in the Y direction that can lead to the flux being deposited onto regions with slightly different pixel sensitivity. We find small drifts ($\lesssim$0.1 pixels) over the first 3 orbits, while the position appears to be more stable over orbits 4 and 5. There are also comparable drifts in the X direction that become evident particularly at the boundary of the water absorption feature where the flux drops rapidly and a shift in wavelength direction will have a big impact. This effect is also visible in the second principal component. Because of the limited impact on our conclusions and the strong undersampling of the PSF, we do not attempt to correct for these drifts. 

\begin{figure}
\epsscale{1}
\plotone{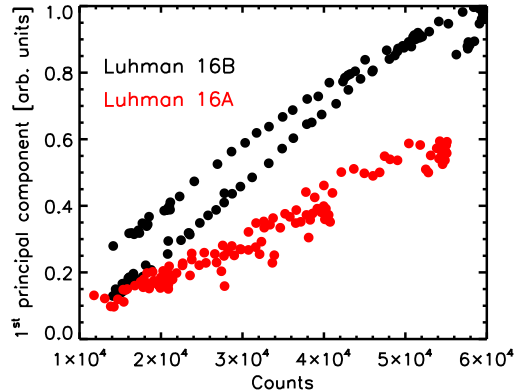}
\caption{Correlation between the average measured counts and the first principal component of the variability (see text) for the Luhman\,16A (red, first orbit only) compared to Luhman\,16B (black). The counts in the first principal component have been scaled arbitrarily. 
\label{fig:var_cnts_correl}}
\end{figure}

We note that the intrinsic spectral variability of brown dwarfs also shows some correlation with the measured flux, as the largest variations are generally measured in the regions of the highest flux \citep{apai13}, while the relative amplitude is lower in the deep water absorption band. However, the relation between measured counts and the counts in the first principal component for Luhman\,16B is distinctly different from the single straight line derived for Luhman\,16A (Fig.\,\ref{fig:var_cnts_correl}). Combined with the absence of any evidence for periodicity, the variations of Luhman\,16A are therefore unlikely to be astrophysical in origin. Nevertheless, we cannot fully exclude that we may have removed small real astrophysical variability for Luhman\,16A that is either aperiodic or of considerably longer period than the observations. 

In Sect.\,\ref{sec:corr} we derive a correction for these errors and conclude that Luhman\,16A is most likely non-variable above $\approx0.2-0.4\%$ between 1.08 and 1.66 $\mu$m. 

We also calculate the influence that the small amount of overlapping flux from the B component could have on A. The region extracted for the A component would correspond to flux from B that is located between 7 and 14 pixels away from its peak. According to the aperture correction values of \citet{kuntschner11}, the amount of flux in this region is 1-1.5\% depending on wavelength. Assuming a relative amplitude of 10\% for B (cf Sect.\,\ref{sec:luh16b}) implies variations of 0.1-0.15\% for A, which is smaller than our remaining errors. 

\begin{figure*}
\epsscale{1.15}
\plotone{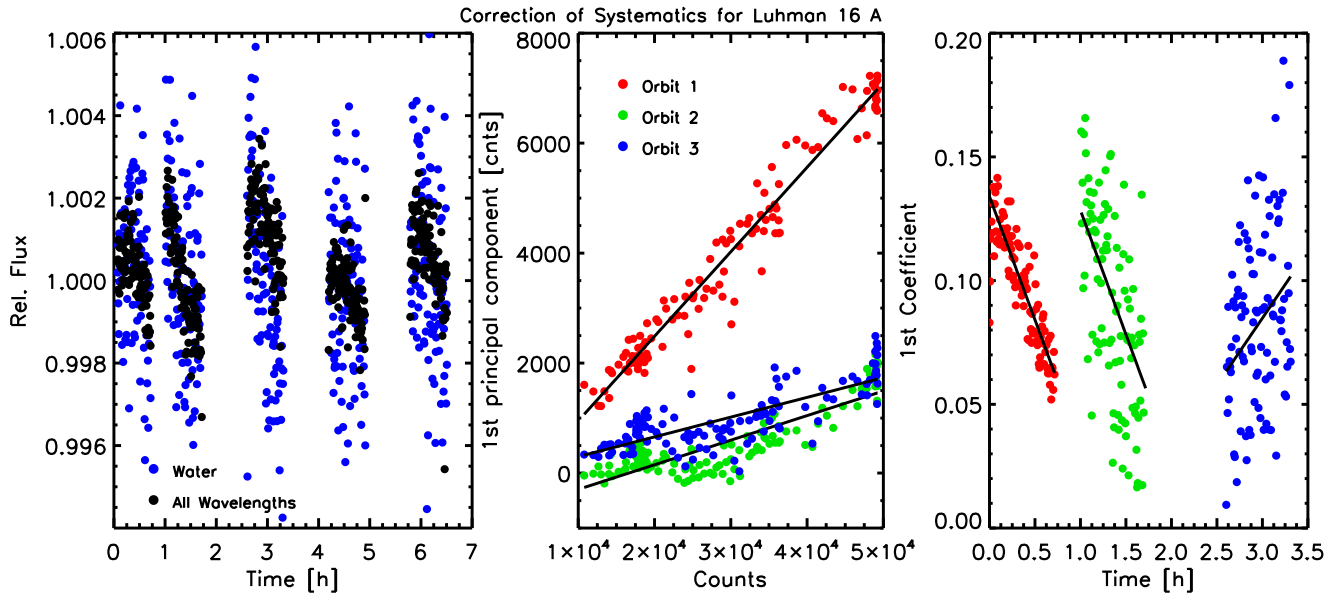}
\caption{Left: Corrected light curve for Luhman\,16 A integrated over all wavelengths (black dots) and only the water absorption band at 1.35-1.44 $\mu$m (blue). Middle: First principal component (see text) as a function of number of counts of the mean flux at a given wavelength for orbits 1 (red), 2 (green) and 3 (blue) with the best corresponding linear fit (black line). The fitted functions are: Orbit 1: $-588+0.154 x$, orbit 2: $-744+0.045 x$, orbit 3: $-60.5+0.036 x$. Right: Coefficients corresponding to the first principal component with the best corresponding linear fit. The fitted functions are: orbit 1: $0.134-0.100 x$, orbit 2: $0.229-0.100  x$, orbit 3: $-0.078+0.054 x$. 
\label{fig:luh16a_corr}}
\end{figure*}

\subsubsection{Correction of systematic errors}
\label{sec:corr}

Because the count rate of Luhman\,16B is very comparable to Luhman\,16A, the systematics in the first orbit in particular may also have a non-negligible impact on the light curves of Luhman\,16B. Here we derive a correction from the principal component analysis of the A component by finding a parametrized form of $a_1(t) e_1(\lambda)$ that is then subtracted from the measured flux $F(\lambda,t)$ for both A and B. For the first three orbits, the parametrization of $a_1(t)$ is determined by a simple linear fit as a function of time (Fig.\,\ref{fig:luh16a_corr}, right), where we disregard the first 10 points in the first orbit that form a separate steep rising ramp. For $e_1(\lambda)$, we find a linear fit as a function of the mean number of counts at a given wavelength (Fig.\,\ref{fig:luh16a_corr}, middle). While this approach is most valid for the first orbit, to first order it also corrects the much smaller variations in the second and third orbits. For B we then use this function and interpolate or extrapolate to the measured counts at each wavelength point. In the left panel of Fig.\,\ref{fig:luh16a_corr} we show that after subtracting the parametrized form of $a_1(t) e_1(\lambda)$ from the flux of the A component, the light curve in the first orbit is constant to within $\approx\pm0.2\%$. We note that within each orbit there remain small trends, perhaps related to small drifts on the detector, that we do not correct for. The standard deviation over the data set of Luhman\,16A as a function of wavelength lies between 0.2\% (flux peaks) and 0.4\% (water absorption band and blue edge of the spectrum). This error cannot be assumed Gaussian and therefore is not significantly smaller for integrated light curves than for individual wavelength points. 

\begin{figure*}
\epsscale{1.1}
\plotone{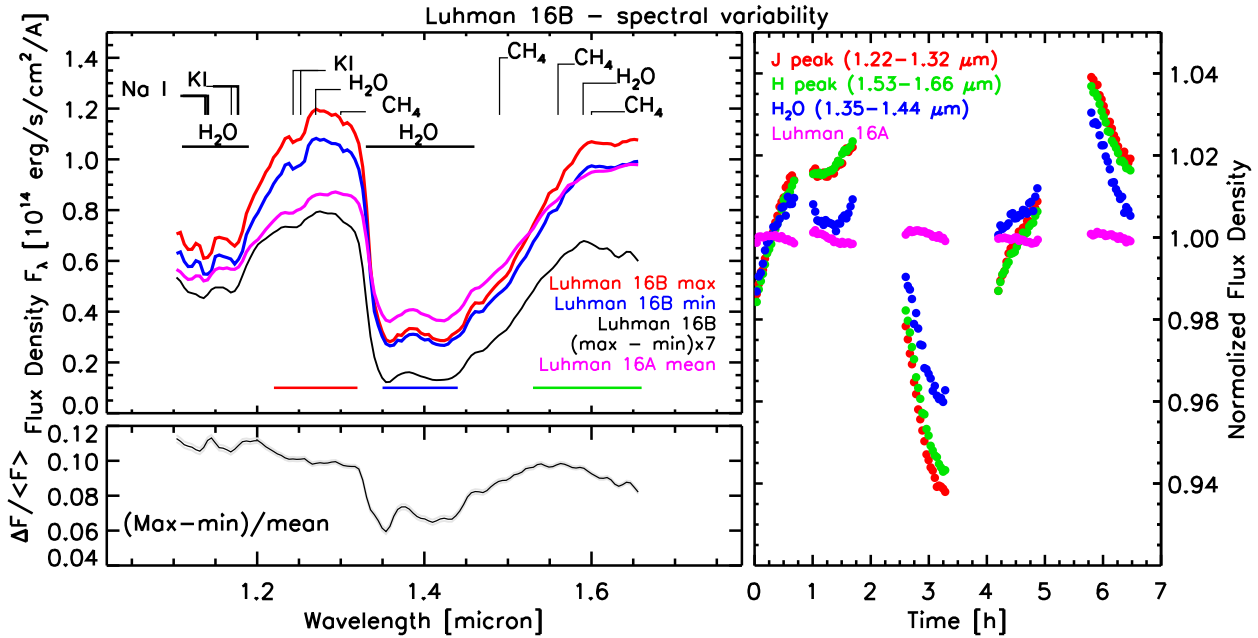}
\caption{Observations of Luhman\,16B. Top Left: Maximum and minimum spectrum compared to the average spectrum of Luhman\,16A. The absolute variability, i.e. the difference between the maximum and minimum is also shown (scaled by a factor of 7 for visibility). The colored horizontal lines indicate the wavelength range over which the spectra were integrated to derive the light curves on the right. Bottom left: Relative amplitude as a function of wavelength, i.e. the difference of the maximum to the minimum divided my the mean. The $1\sigma$ error is shown as a gray band. Right: Integrated light curves of Luhman\,16B derived by integrating the counts in different spectral regions. The integrated light curve over the whole wavelength range for Luhman\,16A is shown for comparison. The small variations correspond to systematic errors. 
\label{fig:luh16b}}
\end{figure*}

\subsection{Luhman\,16B}
\label{sec:luh16b}

Luhman\,16B is variable at all observed wavelengths at a level several times larger than the systematics observed for Luhman\,16A (cf Sect.\,\ref{sec:luh16a}). We apply the derived instrumental correction (cf. Sect: \ref{sec:corr}) to the data of Luhman\,16B and proceed with the corrected spectra. However, we also performed the subsequent analysis on the uncorrected data for comparison, as well as on data corrected by simply dividing by the normalized raw light curves of Luhman\,16A at all wavelengths. The only notable differences lie in the shape of the light curves within the first two orbits. However, the basic conclusions do not depend on whether we correct the instrument systematics or not. 

We removed the first five measurements where a steep ramp was found in the data for Luhman\,16A, as well as 7 points that were strong outliers. We then binned the data to a cadence of 2.14 min, typically averaging 5 observations except where points had been removed. 

Figure\,\ref{fig:luh16b} presents a summary of the observations of Luhman\,16B. It shows the maximum and minimum spectrum (each averaged over 3 binned or 15 individual spectra) compared to the mean spectrum of Luhman\,16A, as well as the relative ampitude as a function of wavelength, i.e. the difference of the maximum to the minimum spectrum divided by the mean spectrum. This difference has been smoothed with a gaussian with an FWHM of 2 pixels, which corresponds to one resolution element. Furthermore, the full light curve integrated over different wavelength regions is also shown: the J band and H band peak and the deep water absorption band at 1.4~$\mu$m. For comparison and to give an estimate of the uncertainties, the fully integrated light curve of Luhman\,16A is also shown. The light curve shape is not sinusoidal at all wavelengths. We find a rapid steep drop in the third orbit (9.5\%/h at the steepest location) in J band. The minimum brightness is likely reached inside the gap shortly after the third orbit. The light curve shows three local maxima with different peak brightness: at the end or just after of the first orbit, after the second orbit, and before the fifth orbit, where the overall maximum is likely reached. The peak-to-valley amplitude lies likely between 10-12\% in the J band peak and is about 1\% lower in the H band peak and 3.5\% lower in the water band. The water band light curve shows small deviations in the shape compared to the  J and H peak light curves, these deviations are discussed in more detail in Sect.\,\ref{sec:pcaB}. 

If the rotation period of Luhman\,16B is $\sim$5~h \citep{gillon13, burgasser14}, the data in orbits 2 and 5 cover almost the same rotational phase. However, significant differences are visible in our observations both in the brightness and shape of the light curves. These differences are discussed in more detail in Sect.\,\ref{sec:evol}. 

The wavelength dependence of the relative amplitude shows a gradual decrease from 11\% at the blue edge of the spectrum (1.1~$\mu$m) to 8\% at the red edge, interrupted by a sharp drop to only about 6\% at 1.35~$\mu$m, where the flux also drops sharply because of a deep water absorption band. The amplitude gradually increases again with wavelength beyond 1.44~$\mu$m to the same level prior to the drop. It is notable that while the absolute variability peaks in the J and H flux peaks, the flux peaks do not correspond to the maximum relative amplitude in J and H, which is found on the increasing slope blueward of the peak. In the absorption band between 1.1 to 1.18~$\mu$m there is also a drop in the ratio, but it is much smaller than in the 1.4~$\mu$m band. An interesting effect is seen in the Na I feature at 1.14 $\mu$m: even though the flux in this feature is lower than at surrounding wavelengths, the relative amplitude is higher. We do not find comparable anti-correlation in any other absorption feature. 

To determine which of the small features in the relative amplitude as a function of wavelength are real, we determine the white noise error per resolution element. Because the flux levels are very comparable between Luhman\,16A and B, this error will also be of similar size. We therefore use the non-variable Luhman\,16A to determine this error to avoid the intrinsic variability of Luhman\,16B to overinflate our errors. Assuming that the non-Gaussian trends in the light curves of Luhman\,16A appear on time scales longer than a few minutes, we calculate the standard deviation between 5 neighboring (unbinned) spectra and then average over all these measurements. The error in each wavelength point for each binned spectrum lies between $\approx0.2\%$ (J and H band peaks), and 0.3\% ( water absorption band and blue spectrum edge). For the relative amplitude $\Delta F/<F>$, we increased the binning and smoothed over one resolution element, the error per resolution element there lies between 0.1 and 0.2\%. This error is marked by the gray band in Figure\,\ref{fig:luh16b}. The bump in the Na I line, as well as the dip at 1.63 $\mu$m appear to be real, as well as much of the structure within the water absorption band. 

\begin{figure*}
\epsscale{1.1}
\plotone{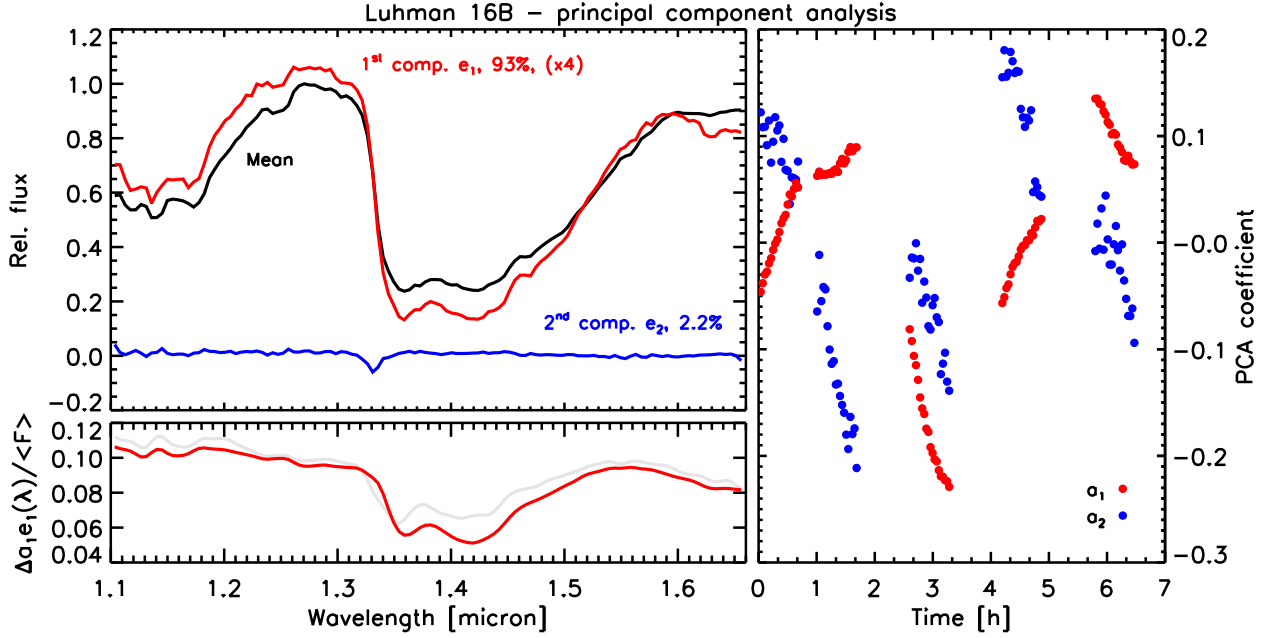}
\caption{Principal component analysis for Luhman\,16B. Top Left: First (red) and second (blue) principal components compared to the mean spectrum (black). They are normalized to the maximum of the mean component. The first component is additionally scaled up by a factor of 4 for better visibility. Bottom left: Variability amplitude (maximum difference of coefficients multiplied by the component and divided by the mean) derived from the first component (red) compared to the variability amplitude from the maximum and minimum spectrum (gray, same as in Fig.\,\ref{fig:luh16b}). Right: Coefficients $c$ of the first and second components. 
\label{fig:pcaB}}
\end{figure*}

\subsubsection{Principal Component Analysis}
\label{sec:pcaB}

\begin{figure}
\epsscale{1.1}
\plotone{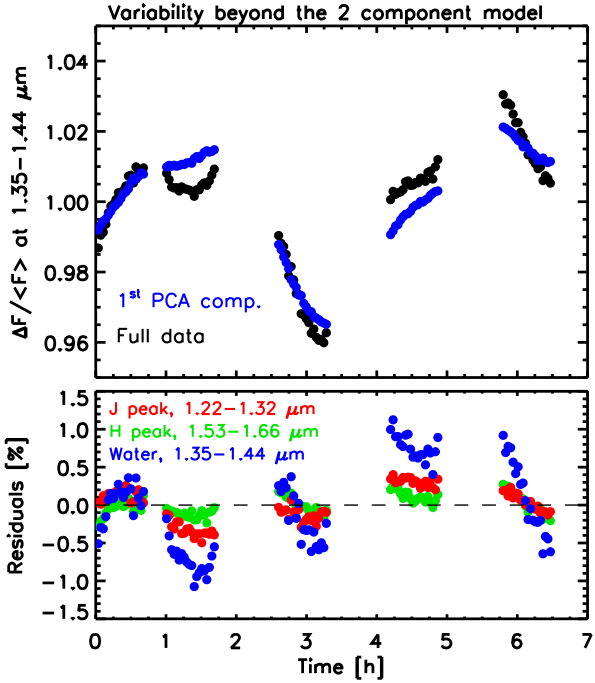}
\caption{Top: Light curves of Luhman\,16B obtained from integrating the spectral time series over the water band between 1.35 and 1.44 $\mu$m. The black points were obtained from the full data cube, while the blue points were obtained from the spectra reconstructed using only the first principal component, i.e. assuming a two-component cloud model. Bottom: Difference between the light curve using the full data and using only the first principal component for the water band, compared to the J band peak (red) and H band peak (green). 
\label{fig:pca_vs_water}}
\end{figure}

Analogous to the analysis done in \citet{apai13} we perform a principal component analysis on the data to determine the number of independent variable components. We subtract the mean spectrum over all orbits and then calculate the components that are variable on top of the mean. We find that 93\% of the variability is described by the first component. The first component and its coefficient, which is essentially identical in shape to the integrated light curve, is shown in Fig.\,\ref{fig:pcaB}. 
We also show the second component which accounts for 2.2\%. However, this component clearly corresponds to an instrumental effect as its coefficient shows small decreasing trends within each orbit similar to what was found for Luhman\,16A. The spectral signature of this component shows a feature at the position of the rapid flux drop into the 1.4 $\mu$m water band and at the spectral edges. This is indicative of positional shifts in x direction during one orbit. All other components account for  $<0.5\%$ of the variability. We can therefore reliably conclude that the spectral variability is for the most part characterized by only two distinct photospheric structures whose visible fractional coverage varies as the brown dwarf rotates. 

\begin{figure*}
\epsscale{1.1}
\plotone{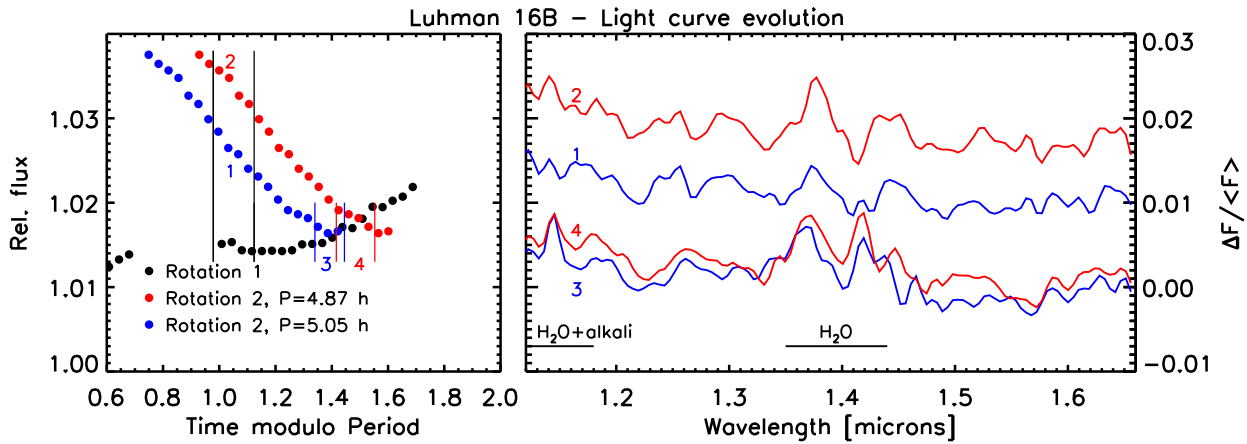}
\caption{Light curve and spectral evolution of Luhman\,16B within one rotation period. Left: Part of the phase folded integrated light curve where the observations covered the same rotational phase twice. The first observation is shown in black, the second is phase folded using two possible periods (4.87 h, red) and (5.05 h, blue). The vertical lines with numbers indicate which part of the light curve was averaged to analyse the spectra. Right: Difference of spectra $(S2 - S1)/((S1 + S2)/2)$ where S1 and S2 denote the spectra taken in the first or second rotational phase at the positions indicated on the left plot. 
\label{fig:evol}}
\end{figure*}

We reconstruct a time series using only the mean and first component spectrum, $F'(t,\lambda) = <F(t,\lambda)>_t + a_1(t) e_1(\lambda)$ to remove some of the noise and instrumental systematics contained in higher order components. In particular, we recalculate the difference between the minimum and maximum spectrum which characterizes the wavelength dependent amplitude changes (Fig.\,\ref{fig:pcaB}, bottom left). In this case, we have used the full time series, and not only the actual maximum and minimum spectrum, to derive this wavelength dependence. This should help average out systematic uncertainties that may be particularly problematic as they are largest between the beginning and end of an orbit. Since the maximum occurs at the beginning of orbit 5 and the minimum of at the end of orbit 3, these systematics may have a significant effect on the ratio shown in Fig.\,\ref{fig:luh16b}. Indeed, we find that $\Delta a_1 e_1(\lambda)/<F>$, where $\Delta a_1$ is the difference between the maximum and minimum of $a_1(t)$, is approximately 0.4\% lower for most wavelengths, and 1-1.5\% in the water band, than $\Delta F/<F>$, which was derived from only the maximum and minimum spectra. This agrees with the derived systematic uncertainties (Sect.\,\ref{sec:corr}). However, the shape of the ratio over the water band is also different, which indicates that the water varies slightly differently over the course of observations. Indeed, comparing a reconstructed light curve with only one PCA component in the water band with the original light curve reveals significant differences (Fig.\,\ref{fig:pca_vs_water}). Because the signal in the water band is much lower and the noise larger than in the other regions, this difference is hidden within a combination of higher order components in the principal component analysis. It indicates that while a two-component model can broadly account for most of the spectral variability, additional smaller variations in the deep water band are also present. The differences are overall larger than the $\approx\pm0.5\%$ uncertainty found in the Luhman\,16A water light curve.  Small differences of up to $\pm 0.3\%$ are also found for the J band peak but it is unclear whether these deviations are of astrophysical origin. 

We don't find any other significant differences in the wavelength dependence, but it is possible that in this procedure we have removed small astrophysical variations that were hidden within the higher order components that we classified as random noise or systematic errors. However, because only two components clearly contribute to most of the variability, we will only attempt to build a two-component model (cf. Sect.\,\ref{sec:model}) and therefore also only fit to the dimensionally reduced data. 

\subsubsection{Evolution over one rotational period}
\label{sec:evol}

The rotation period of Luhman\,16B has been estimated to be $4.87\pm0.01$~h by \citet{gillon13} and $5.05\pm0.1$~h by \citet{burgasser14} based on observations spanning several days. From our observations over 6.5 h we can therefore discuss the spectral changes that have happened over the course of one rotation. In Fig.\,\ref{fig:evol}, we show the parts of the phase-folded light curve where we have coverage for two rotations. Clearly, the light curve has evolved significantly over only one rotation period. At the beginning of the overlap, we find a higher brightness of 1-2\% for the second rotation than for the first, and presumably an even larger difference during the gap where observations are missing for the first rotation. Also, in the first rotation, the light curve is increasing, but it is rapidly decreasing at the same phase during the second rotation. 

The spectra between the two rotational phases are largely identical within the noise level (Fig.\,\ref{fig:evol}, right), except for an offset where the brightness is different. The only difference is found within the deep water band at 1.4~$\mu$m, as well as in the absorption band shortward of 1.2~$\mu$m. In the second rotational phase there is slightly more flux ($\sim0.5\%)$ in the water band compared to the other wavelengths. This is interesting because from our previous analysis we would expect the relative flux in the water band to be slightly lower in the brighter rotational phase because the overall variability amplitude is lower in the water band. The fact that we find the opposite provides additional evidence that there are additional small variations present in the water band that are not captured by our principal component analysis. 

\subsection{Relative angular distance and position angle }

Luhman\,16A and B are the only two sources with significant signal in the field of view of our 256$\times$256 pixels sub-array WFC3/IR images. Therefore, we cannot obtain their positions with respect to ``fixed" background objects. The absolute angular distance and positional angle between Luhman\,16A and B will ultimately be at the mercy of the instrument calibration and characterization of its stability.
 
To measure the positions in individual images we employed the PSF-fit software and library developed by Jay Anderson\footnote{Available at
\texttt{http://www.stsci.edu/\textasciitilde jayander/WFC3/WFC3IR\_PSFs/}}. It is an adaptation of the program first presented in \citet{anderson06}; a small additional adaptation was done by us to read 256$\times$256 subarray images. With only the two brown dwarfs visible, we cannot meaningfully solve for both the positions and PSF of the objects, we therefore adapted a library PSF. We used the PSF for filter F139M, which is the closest available to that of the filter F132N in which our direct images were taken. We extracted the raw pixel coordinates from the \texttt{flt} images using the \texttt{img2xym\_wfc3ir} routine and corrected them for the geometric distortion with the program \texttt{wfc3ir\_cg}, which uses the best available solution currently available at the Space Telescope Science Institute. 

We then computed, for each of the two axes, the difference in geometric-distortion corrected positions between Luhman\,16A and B. The average and scatter of their relative angular distance in pixels from our five images is $ \Delta r = 10.292 \pm 0.019$ pixels. Adopting a pixel scale of 120.99~mas pixel$^{-1}$ for the geometric distortion correction developed by Anderson, the separation of the two components of Luhman\,16 amounts to $1245.2\pm2.3$~mas. The uncertainty reflects only the internal errors and likely underestimates the true errors which might cancel out in this particular data set where images were collected under (nominally) identical conditions. The WFC3/IR pixel scale is stable down to 5 parts in 10\,000 (Jay Anderson, private communication), and it is significantly more stable than for WFC3/UVIS. Therefore we add to the above quoted uncertainty an additional error of $\pm 1.3$ mas. 

The above method is the most accurate to derive relative separations, but it is not normalized to the guide stars on which the roll angle relies. To derive the position angle, we therefore use the \texttt{drz} images output from the HST pipeline, which are distortion corrected and include WCS information, however no library PSFs are available. Our five original direct images were combined by the pipeline into two separate distortion corrected images. We measure the positions of the binary components by determining the photo-centroid. We find a position angle $PA = 311\pm3^{\circ}$. The dominating error component derives from the uncertainty in the photo-centroid determination. We conservatively assume 0.5 pixels error in the distances $\Delta x$ and $\Delta y$ between the binary. The additional uncertainty in the north direction because of the error in roll angle is less than an arc minute, and therefore negligible. 

The separation of Luhman\,16AB, here measured as 1\farcs25, has clearly decreased since the epoch where its binarity was first recognized. The binary separation was given as 1\farcs5 by \citet{luhman13} from a GMOS measurement taken on February 23, 2013, although without specifying an error bar. Immediate follow up by \citet{burgasser13} indicated a separation of $1\farcs54\pm0\farcs04$ on March 12, 2013 at a position angle of $313\pm3^{\circ}$.  
Further astrometric measurements were obtained by \citet{boffin14} between April 14 and June 22, 2013, where the separation decreased from $\approx1\farcs43$ to 1\farcs36. Within 8 months, the projected binary separation has therefore shrunk by approximately 0\farcs3. The position angle has remained constant within error bars. We refrain from constraining the binary orbit because dedicated programs to obtain high-precision astrometry of the system with both HST and FORS2 are ongoing that will deliver much more accurate results. However, our result shows that it will be important for future observing campaigns of this binary to take into account that the separation may have further decreased and that resolving them spatially may become considerably more challenging, especially without adaptive optics from the ground. 

\section{Atmosphere modeling}
\label{sec:model}

We use models based on the \citet{ackerman01} cloud models which have been updated with new opacities for H$_2$ collision-induced absorption (CIA), NH$_3$ and FeH \citep{saumon12}. The models parametrize the clouds with the sedimentation efficiency $f_{\mathrm{sed}}$. A smaller $f_{\mathrm{sed}}$ value corresponds to a vertically and optically thicker cloud. Our analysis uses a grid of models covering $f_{\mathrm{sed}} = 1$, 2, 3 and non-cloudy (nc), effective temperatures $T_{\mathrm{eff}}$ between 900 and 1500 K in 100 K steps, gravity $\log{g} = 4.5$ and 5, and the vertical mixing parameter $K_{zz}$ = 0 and 10$^4$\,cm$^2$s$^{-1}$, which controls departures from equilibrium chemistry \citep[see][for detail]{stephens09}. Because we also fit for the absolute flux level, the radius $R$ of the brown dwarf is an additional free parameter. We use a distance of 2.02 pc determined from the parallax by \citep{boffin14}.

We simultaneously try to match the overall average HST spectrum and the broader 0.8 - 2.5 microns \citep[FIRE spectrum from][]{burgasser13}, as well as the relative amplitude as a function of wavelength $\Delta F/<F>$ for Luhman\,16B. We note that the ground-based FIRE spectrum is strongly affected by telluric water lines between 1.35 and 1.45\,$\mu$m and between 1.8 and 2\,$\mu$m; we disregard the FIRE observations for these regions but we fit the 1.35-1.45\,$\mu$m region from the HST spectrum.

\begin{figure}
\epsscale{1.2}
\plotone{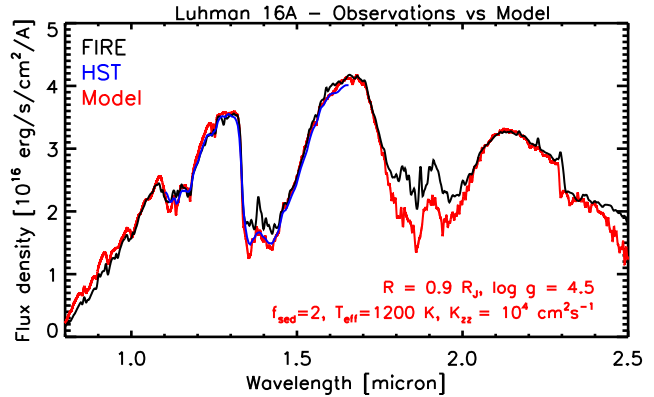}
\caption{HST and FIRE spectra of Luhman\,16A compared to the best-fit model. FIRE spectrum from \citet{burgasser13}. 
\label{fig:model16A}}
\end{figure}

For both Luhman\,16A and B, it is obvious that models with $K_{zz} = 0$\,cm$^2$s$^{-1}$ cannot properly reproduce the shape of the K band spectrum. If we allow for out-of-equilibrium chemistry by setting $K_{zz} = 10^4$ in the calculation of the synthetic spectra, the shape of the K band spectrum can be matched reasonably well. We therefore continue only with those models. We note that the calculation of vertical mixing is also not fully self-consistent because there is no feedback between the modified chemistry and the temperature-pressure profile of the atmosphere based on equilibrium chemistry.  

We first model the spectrum of Luhman\,16A. The only model that provides a very good fit has parameters $T_{\mathrm{eff}}$ = 1200 K, $f_{\mathrm{sed}}$ = 2, $\log{g} = 4.5$, $K_{zz} = 10^4$\,cm$^2$s$^{-1}$ and $R = 0.95$\,$R_{\mathrm{J}}$ (Fig.\,\ref{fig:model16A}). This is about 20\% lower than the evolution radius for these parameters \citep{saumon08}. 

For Luhman\,16B, none of the models reproduce the average spectrum. This is unsurprising, as the variability indicates that we require more than a homogeneous cloud model. Because most of the observed spectral variability can be explained by two components, we attempt to fit a model that is a linear combination of two models with different cloud properties and effective temperatures. We note that because these two models have different temperature-pressure profiles, which would be coupled in reality, the calculations are not self-consistent. First attempts at more realistic patchy cloud models have been made by \citet{marley10} and \citet{morley14}, but that modeling framework is not yet fully ready for large-scale fitting of observations. We defer such patchy cloud modeling to a future paper. Nevertheless, the linear combination models can be seen as a useful first step in constraining the parameter space of possible cloud properties. 

The two models that are linearly combined can have different effective temperatures $T_{\mathrm{eff,1}}$ and $T_{\mathrm{eff,2}}$ and different cloud parameters $f_{\mathrm{sed,1}}$ and $f_{\mathrm{sed,2}}$. $\log{g}$ and $K_{zz}$ have to be equal for the two models. The weights in the linear combination are set using the covering fraction $c_1$ of the warmer model on the hemisphere where the emitted flux is maximal, and the change in covering fraction $\Delta c_1$. Then, the combined flux $F_{\mathrm{max}} = c_1 F_1 + (1 - c_1) F_2$ and  $F_{\mathrm{min}} = (c_1-\Delta c_1) F_1 + (1 - (c_1-\Delta c_1)) F_2$. 

Because the parameter space is large and results of chi-square fitting strongly depend on the weights assigned to different spectral regions, we iteratively fit the models by eye. We only discuss approximate best fits and degeneracies, and defer a more quantitative model fitting to a later paper, once self-consistent patchy cloud models become available. 

\begin{figure}
\epsscale{1.2}
\plotone{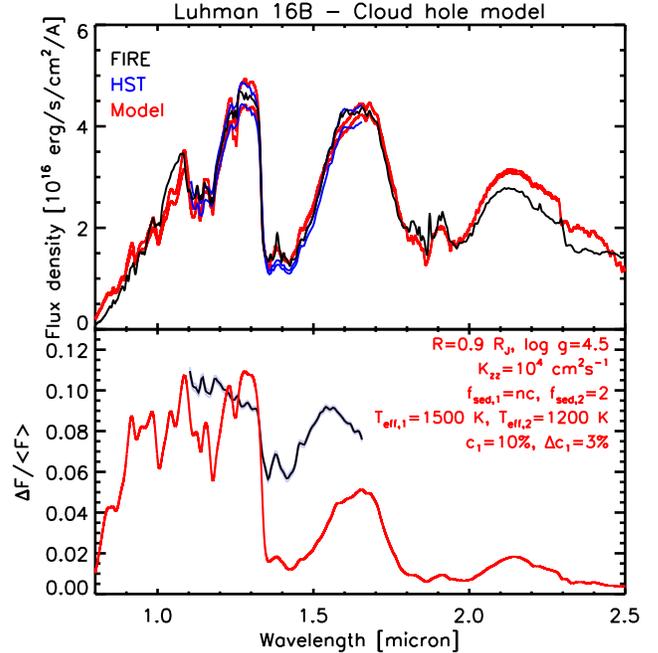}
\caption{Top: Average FIRE spectrum (black) and HST maximum and minimum spectra (blue) of Luhman\,16B compared to a model that introduces a small fraction of hot, non-cloudy regions. Bottom: Relative amplitude as a function of wavelength, i.e. the difference of the maximum to the minimum spectrum divided by the mean for the HST observations compared to the model. Such a model cannot reproduce the observed variability. 
\label{fig:badmodels}}
\end{figure}

We first try cases that use the cloud model from Luhman\,16A as a base ($T_{\mathrm{eff}} = 1200 K$, $\log{g} = 4.5$, $f_{\mathrm{sed}} = 2$, $K_{zz}=10^4$\,cm$^2$s$^{-1}$), and introducing non-cloudy areas to approximate cloud clearings. We find that the spectral shape can be approximately matched - although not perfectly in the K band - by introducing non-cloudy sections with $T_{\mathrm{eff}} = 1500$\,K and covering fractions of 7-10\%. However, the wavelength dependence of the relative amplitudes induced by such cloud holes is very different than observed (Fig.\,\ref{fig:badmodels}). This is unsurprising considering that similar conclusions had already been drawn about the other two known variable early T dwarfs in \citet{radigan12} and \citet{apai13}. Combinations of different cloudy and non-cloudy models do not provide better results. We can therefore exclude holes without cloud opacity as the cause of variability for Luhman\,16B. 

\begin{figure*}
\epsscale{1.2}
\plotone{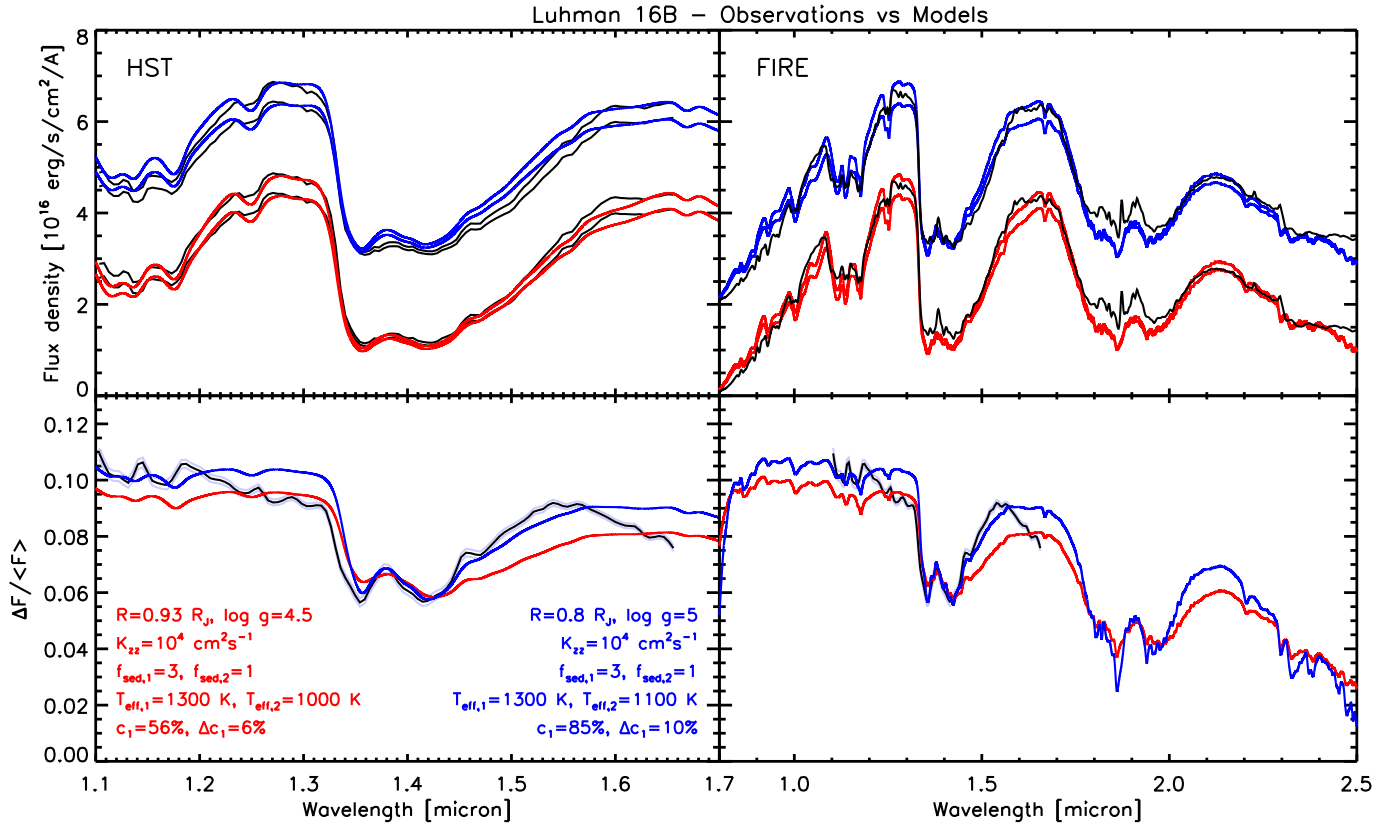}
\caption{Modeling for Luhman\,16B. Top left: HST maximum and minimum spectra (black) and two best-fit models with log g = 4.5 (red) and log g = 5 (blue). For clarity, observation and model for one of the cases are plotted with a vertical offset of $2\cdot 10^{16}$ erg/s/cm/A. Bottom left: Relative amplitude as a function of wavelength, i.e. the difference of the maximum to the minimum divided my the mean. Top right: The same models compared to an average FIRE spectrum covering a wider wavelength range. Bottom right: Model predictions for the relative amplitude for the extended wavelength range. The measured variability from HST is overplotted. 
\label{fig:model}}
\end{figure*}

We proceed with combining models with different values of $f_{\mathrm{sed}}$. We cannot conclusively discriminate between models of different gravities, but the best-fit cloud parameters change somewhat depending on the choice of gravity. We therefore provide best-fit models for both log g = 4.5 and 5. These are shown in Fig.\,\ref{fig:model}. Regardless of log g, we find that we require a model combination of $f_{\mathrm{sed,1}} = 3$ and $f_{\mathrm{sed,2}} = 1$ and effective temperatures in the 1000 - 1400 K range, with the thinner cloud (larger $f_{\mathrm{sed}}$) being warmer.  

Models with other combinations of $f_{\mathrm{sed}}$ cannot fit the average spectrum and/or show strong deviations in the fit of the relative amplitudes. The effective temperature of the two models is degenerate with the covering fraction, but we can restrict $T_{\mathrm{eff,1}}-T_{\mathrm{eff,2}}$ to 200-300 K because lower and higher temperature differences fail to give the correct relative amplitude in the water band compared to the flux peaks. For both $\log{g}$ values, the best models have $T_{\mathrm{eff,1}} = 1300$\,K, while the cooler model has $T_{\mathrm{eff,2}} = 1000$\,K for $\log{g} = 4.5$ and $T_{\mathrm{eff,2}} = 1100$\,K for $\log{g} = 5$. Although we cannot fully exclude models with temperatures of 100\,K higher or lower, the spectral fit tends to be worse especially in the K band. 

In general, the smaller the temperature difference, and the lower the effective temperature of the warmer model, the larger the covering fraction $c_1$ of the warmer cloud has to be. The covering fraction of the warmer cloud can lie between $\approx$30 and 90\%. For our best-fit models, we find $c_1 = 56\%$ and $\Delta c_1 = 6\%$ in the case of $\log{g} = 4.5$, but $c_{1} = 85\%$ and $\Delta c_1 = 10\%$ for $\log{g} =  5$, where $\Delta c_1$ is the change in the covering fraction of model 1 between the two hemispheres, as outline above. Derived radii are physically reasonable (0.93 R$_J$ or 0.8 R$_J$) but remain systematically lower than those corresponding to the evolution radius \citep{saumon08}. We note that these results could potentially change, in particular with respect to the covering fractions, if we also allowed intermediate values for $T_{\mathrm{eff}}$ and/or $f_{\mathrm{sed}}$, or values of $f_{\mathrm{sed}}$ below 1. 

The overall spectral shape fits surprisingly well, although the flux in the water band and wing is overestimated in the $\log{g} = 5$ case. On the other hand, for $\log{g} = 4.5$ the shape of the H band peak in the FIRE spectrum is not well fit. The main characteristics of the spectral variability are also reproduced fairly well, in particular the approximate relative amplitude in the J, water and H bands. We cannot find any model that accurately fits the slope between 1.1 and 1.3 $\mu$m, nor wavelength dependence of the relative amplitude across the H band beyond 1.56 $\mu$m. However, the amplitude difference between the absorption bands at 1.14 and 1.4 $\mu$m and the outside continuum are approximately reproduced.  The differences in the relative amplitude between model and observations are at most 2\%. The prediction that the K band amplitude is significantly lower is consistent with the results in \citet{burgasser14}. 

\section{Discussion}

The formation and growth of cloud holes through the L/T transition has been suggested as a theory to simultaneously explain the color evolution, the re-emergence of the FeH 0.9896$\,\mu$m Wing-Ford band, and the occurrence of photometric variability \citep{ackerman01, burgasser02, marley10}. However, recent observational results have already revealed a more complex picture: cloud holes could not explain the spectroscopic variability of 2M2139 and SIMP0136 \citep{apai13, buenzli14}, and the FeH band is equally strong in Luhman\,16A and B \citep{faherty14}. Our spectroscopic variability observations of Luhman\,16A and B strongly support these findings and show that the cloud evolution through the L/T transition is significantly more complicated than simple formation of cloud holes. Comparing the single cloud layer found for Luhman\,16A with the two for Luhman\,16B, our best-fit model suggests both different effective temperature and sedimentation efficiency. We can also confidently exclude full cloud-clearings, i.e. regions without cloud opacity, although the thin patchy clouds within a background of thick clouds can still be regarded as holes in the sense that more bright flux will emerge from these regions. However, our results suggest that the whole cloud cover can be significantly different from late L to early T spectral type, as opposed to having a similar cloud layer that is simply broken up or thinned out at the beginning of the L/T transition. Perhaps surprisingly, we find the thick cloud component in the T dwarf to be thicker than that of the overall cloud cover on the L dwarf. 

A shortcoming of our model is that the equal depth of the FeH band is not properly reproduced. For Luhman\,16B, the FeH band is predicted to be deeper than for Luhman\,16A, even without the opening of cloud holes. Whether this is an issue related to the FeH opacities or Fe condensation chemistry, the patchy cloud modeling which is not self-consistent, or a more fundamental shortcoming of the cloud model remains unclear. 

With the effective temperature difference between the two cloud layers estimated to 200-300\,K, we arrive at a difference in covering fraction of about 5-10\% between the two hemispheres. The simplest explanation for the rapid light curve evolution is then that the covering fraction changes, perhaps due to rapid redistribution of the clouds. An additional 1-2\% change in the covering fraction can already alter the relative amplitude by a few percent. Periods without or with only very low variability \citep{gillon13} may simply be times when the covering fraction is approximately equal at all longitudes. The additional spectral changes beyond the two-component model suggest that small alterations in the cloud structure beyond the covering fraction can also happen on short time scales, although the major characteristics of the spectroscopic variability are preserved within a single rotation. Further epochs would be required to trace larger scale changes in the cloud structure. Some intriguing differences between our measurements and multi-wavelength photometry taken at an earlier epoch are discussed below.  

\subsection{Comparison to other observations of Luhman\,16}
\label{sec:otherobs}

Since its discovery in 2013, Luhman\,16 has been extensively monitored for time variability from the ground over a broad wavelength range with different instruments. Monitoring over 12 nights with the TRAPPIST instrument through a special I+z filter (750 - 1100+ nm, effective wavelength 910 nm) revealed rapidly changing variability amplitudes between 2-11\% \citep{gillon13}. These observations did not spatially resolve the binary. Assuming that all variability stems from Luhman\,16B, this would translate to effective peak-to-valley amplitudes of 4-22\%. If the amplitudes don't significantly change at wavelengths shortward of our observations, then our observed amplitudes of $\sim10\%$ would fall into an intermediate regime. 

Because of the rapid changes, single filter photometry does not provide useful information on the cloud structure for this object. Simultaneous multi-color photometry was obtained with the GROND instrument by \citet{biller13} on two nights in 6 filters spanning from the optical to the near-infrared (r', i', z', J, H, K). On one night, the observations were made in spatially unresolved mode, in the other in spatially resolved mode. The results in the near-IR from the unresolved observations agree relatively well with our observations: the peak-to-valley amplitude in J band was found to be 14\% in J and 8\% in H band assuming all the variability stems from Luhman\,16B. The $\Delta(J-H)$ color change is larger than in our observations, but we do not fully cover the H band where we would expect the amplitude to keep decreasing. The light curve shape in the J, H and z' bands were very similar, but interestingly, the light curves in r' and i' bands were found to be anti-correlated to J, H and z'. We do not find any anti-correlation in our observations, but our observations do not span the wavelengths where anti-correlation was found. 

The most puzzling set of observations are the spatially resolved GROND observations taken a few days after the unresolved observations. Most interestingly, no variability was found in J band above the photometric uncertainty of 3\%, while the z' and H bands showed clear variations of 9\% in z and 13\% in H band. These observations suggest that at that moment the wavelength dependence of the amplitude was completely different than during the unresolved observations and during our HST observations, as well as compared to the other known variable early T dwarfs (cf. Sect.\,\ref{sec:others}). The different color change in z'-H might be partially explained by the large uncertainty of $\sim4\%$ in the resolved GROND H band observations. However, none of our models could produce significant variability in H and z' bands while simultaneously not having any or only very low variability in J band. Multi-epoch multi-wavelength monitoring would be required to confirm this very unusual change in the color signature of the variability. 

The GROND observations also contained the first K band photometry, which appeared to show an intermediate phase between the phase of the correlated z' and H and anti-correlated r' and i' light curves. \citet{biller13} suggested a similar dependence of the phase on the atmospheric pressure probed in a band as found for the mid-T dwarf 2MASSJ2228-43 \citep{buenzli12}. Our observations do not show any phase shifts, even though the deep water band in our observations would probe pressures lower than the r', i' and K bands. If the K band phase shift is not an artefact of instrumental systematics, the three dimensional structure of the atmosphere of Luhman\,16 is likely to be changing as well. 

The first spectroscopic variability study for Luhman\,16B was performed by \citet{burgasser14} with the SPEX instrument. They obtained a 45 min spectral time series spanning from 0.7 to 2.5~$\mu$m at a resolution of $\lambda/\Delta \lambda \approx120$, together with TRAPPIST photometry spanning 7 h. These observations only looked at relative changes between Luhman\,16A and B, but our observations show that at least in the near-IR the assumption that any variability of Luhman\,16A is negligible compared to B appears to hold. They found decreasing flux of about 10\%/h in Y and J bands and 7\%/h in H band. This is in very good agreement with the color variability in our observations, as well as the maximum flux decrease in our J band observations. They also found some evidence of lower variability in the water band at 1.4 $\mu$m and a continuing decrease into the K band. However, the signal-to-noise ratio in these wavelength regions makes the measured variability non-significant. Here we clearly confirm the lower relative amplitude in the 1.4 $\mu$m water absorption band, and our model fit is fully consistent with the lower amplitudes in K band ($\approx 4\%/h$). 

Our HST observations are overall very similar in the characteristics to the SPEX variability observations, while at least one epoch of GROND multi-wavelength photometry offers a very different picture. Additional epochs of multi-wavelength observations are required to obtain a better understanding of the rapid changes occurring in the atmosphere of Luhman\,16B. 

Comparing our observations to the surface map derived with Doppler imaging \citep{crossfield14} is difficult, because it is unclear how variations in the molecular line shape relate to the flux variability we observe with HST. Furthermore, the method of reconstruction for the surface map yields a smooth brightness distribution which does not easily translate into a two-component coverage fraction. It is possible that the largest dark spot found with Doppler imaging may correspond to the region where the coverage fraction of the thicker clouds is largest. 

\subsection{Comparison to other variable brown dwarfs}
\label{sec:others}

Luhman\,16B is the fourth variable brown dwarf for which we have obtained HST spectroscopic time series covering one or more rotation periods after the T6.5 dwarf 2M2228-43 \citep{buenzli12} and the two early T dwarfs 2M2139 and SIMP0136 \citep{apai13}. Already from this very small sample a trend in variability characteristics emerged as a function of spectral type. The two early T dwarfs showed very similar spectral dependence of the variability: the light curves were in phase at all wavelengths and with only slightly larger amplitude in J than H band, but significantly lower amplitude in the deep water absorption band. The mid-T, on the other hand, showed complex phase shifts, the largest variability in the water band and a larger amplitude in H than J band. 

The characteristics of the spectral variability of Luhman\,16 is again very similar to that of the two other early T dwarfs: all variability is in phase (excluding the anti-correlation observed with GROND, for which we do not have similar information for the other two objects), and the water band variability is lower than in the J and H bands. 

\begin{figure}
\epsscale{1.15}
\plotone{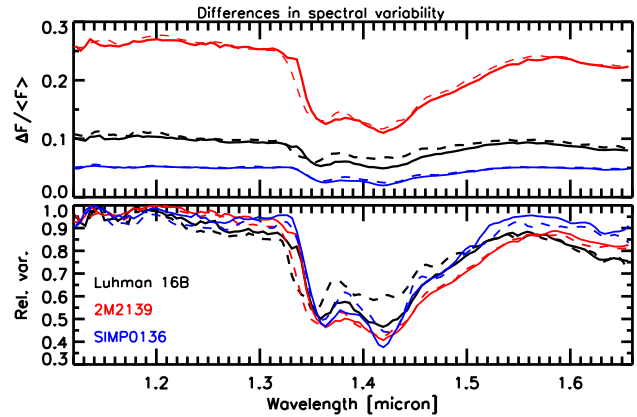}
\caption{Wavelength dependence of the variability for Luhman\,16B (black), 2M2139 (red) and SIMP0136 (blue). Top: variability amplitude as a function of wavelength calculated as the difference between the maximum and minimum spectrum divided by the mean. Bottom: The same but normalized to the maximum amplitude. The dashed lines directly use the observed maximum and minimum spectra averaged over a few minutes, the solid lines use the maximum and minimum spectra reconstructed by using only the mean and first principal component. 
\label{fig:compare}}
\end{figure}

In Figure\,\ref{fig:compare} we directly compare the spectral variability for the three objects both in absolute terms and scaled to the same maximum amplitude. We derived the variability difference in two ways: directly from the maximum and minimum observed spectra (averaged over a few minutes to improve the S/N), as well as from the reconstructed maximum and minimum spectra using only the mean and first component from the principal component analysis. For 2M2139 and SIMP0136 we performed a principal component analysis exactly analogous to Luhman\,16B. For 2M2139 and SIMP0136 the spectral variability is very similar in both cases, unlike for Luhman\,16B where we found a significant difference in the water band, as discussed in Sect.\,\ref{sec:pcaB}. 

Comparing only the variability derived from the principal component analysis, i.e. corresponding to a simple two-component model, we find that except for the different amplitudes, the spectral signature of the variability is remarkably similar for all three objects. The relative amplitude in the water band is always approximately half of the maximum relative amplitude, which is always found at 1.14 $\mu$m. All three objects also share very similar small dips at 1.12 and 1.16 $\mu$m. A small but significant difference is primarily found in the overall slope from the J to the H band: for SIMP0136, the latest type object, we find the flattest slope with only a 10\% decrease from 1.2 to 1.66 $\mu$m. The slope is intermediate for 2M2139 with an 18\% decrease, while the steepest slope is found for the earliest spectral type object Luhman\,16B. There, the slope is already notable across the J band, and we find a total of 25\% decrease from 1.2 to 1.66 $\mu$m. These values are consistent with the J-H variability found in different epochs for Luhman\,16B \citep{burgasser14} and 2M2139 \citep{radigan12}, although we note that for 2M2139 significant changes in J-K color were found during nights only a few days apart. 

In terms of variability amplitudes, Luhman\,16B (10.5\%) is here found to be intermediate between 2M2139 (27\%) and SIMP0136 (5.5\%), although these values are known to change significantly between epochs. Luhman\,16B is also intermediate in terms of rotation period (5 h), compared to 7.8 h for 2M2139 and 2.4 h for SIMP0136. This was already noted by \citet{burgasser14} who discuss the connection to the Rhines scale, which provides an estimate of the maximum possible feature size and scales linearly with rotation period and wind speed \citep[see also][]{apai13}. For a more slowly rotating object, the occurrence of larger features are possible. At equal covering fraction, larger features would also imply fewer features, which then have a higher probability of being distributed unequally. For these three objects, the results are consistent with potentially larger features on the more slowly rotating objects. However, without good constraints on the covering fractions of the respective components, no strong conclusions can be drawn. 

\section{Conclusions}

As our nearest brown dwarf neighbors, the Luhman\,16AB system offers unprecedented opportunities to further our understanding of the L/T transition. We have obtained very high signal-to-noise, spatially resolved spectroscopic time series of both objects with HST that reveal a highly patchy cloud structure for Luhman\,16B.  

Our main conclusions are:

\begin{itemize}

\item{Luhman\,16A is not variable at 1.1-1.66 $\mu$m above our instrumental systematics of $\approx$0.2-0.4\% depending on wavelength. A model with a single, homogeneous cloud layer with T$_{\mathrm{eff}}$ = 1200 K, $f_{\mathrm{sed}} = 2$, $\log{g}=4.5$ and out-of-equilibrium chemistry with $K_{zz} = 10^4$\,cm$^2$s$^{-1}$ provides a good fit to the spectrum.}

\item{Luhman\,16B varies with a maximum peak-to-peak amplitude of more than 10\% in the near-IR, but the relative amplitude is lower in the deep water absorption band at 1.4 $\mu$m. Outside the absorption band there is also a small decrease in the relative amplitude with increasing wavelength.}

\item{A two-component cloud model can explain most, but not all, of the variability of Luhman 16\,B, with the primary differences seen in the water absorption band. A combination of a warmer, thinner cloud with  T$_{\mathrm{eff}} = 1300$\,K and $f_{\mathrm{sed}} = 3$ and a cooler, thicker cloud with T$_{\mathrm{eff}} = 1000-1100$\ K and $f_{\mathrm{sed}} = 1$ provide a decent fit to both the spectrum and relative amplitude as a function of wavelength, but the covering fraction of each cloud is degenerate with other model parameters. The relative covering fraction varies by 5-10\% between the two hemispheres. Out-of-equilibrium chemistry ($K_{zz} = 10^4$\,cm$^2$s$^{-1}$) is also required to fit the spectrum.} 

\item{We can firmly exclude the existence of areas without cloud opacity, suggesting that the primary difference in the color and the variability between a co-eval late L and early T dwarf is not caused by the opening and clearing of deep holes. Instead, we find a difference in the cloud structure before the L/T transition (overall intermediate cloud thickness) to within the transition (patchy coverage with thicker and thinner clouds).}

\item{The relative amplitude shows very similar wavelength dependent behavior for the three known highly variable early T dwarfs, suggesting similar underlying cloud structure that may be typical for these objects.}

\item{The light curve shape and amplitude of Luhman\,16B evolves significantly on the time scale of only one rotation period. Between epochs, the wavelength dependence of the variability may also change significantly, although it is unclear what mechanism could cause these changes.}

\item{The projected separation of Luhman\,16A and B has decreased by 0\farcs3 within a time-span of 8 months.}

\end{itemize}

Further monitoring of the orbital evolution with HST (Program \#13748), Gaia, and from the ground will eventually result in a dynamical mass measurement, a crucial input for the constraint of atmospheric models. Combined with continuing monitoring of the variability of Luhman\,16B across a wide wavelength range and the direct comparison to its companion Luhman\,16A, it will be possible to put together a comprehensive model of its atmosphere and weather patterns that cannot be obtained for any other substellar object. 

\acknowledgements
We thank the staff at Space Telescope Science Institute (STScI), in particular Amber Armstrong, for the coordination and scheduling of the observations. We also thank Jay Anderson for providing information and routines regarding the distortion correction for WFC3. We thank Adam Burgasser, Alexei Kniazev and Jackie Faherty for providing their published spectra of Luhman\,16AB in electronic form. Based on observations made with the NASA/ESA Hubble Space Telescope, obtained at the Space Telescope Science Institute, which is operated by the Association of Universities for Research in Astronomy, Inc., under NASA contract NAS 5-26555. These observations are associated with program \# 13280. 
Support for program \#13280 was provided by NASA through a grant from the Space Telescope Science Institute. EB was supported by the Swiss National Science Foundation (SNSF). This research has made use of the SIMBAD database, operated at CDS, Strasbourg, France, and of NASA's Astrophysics Data System Bibliographic Services.

\end{document}